\documentstyle[12pt,psfig,aaspp4]{article}

\begin{document}

\title{Using Perturbative Least Action to Reconstruct the Local Group}

\author{David M. Goldberg}
\authoremail{goldberg@yale.princeton.edu}
\affil{Yale University, Astronomy Dept.,  New Haven, CT, 06520-8101}
\affil{Princeton University Observatory, Princeton, NJ 08544-1001}

\begin{abstract}
In this paper, we apply the Perturbative Least Action Method to model
the Local Group of galaxies for various Cosmological models.  We show
that though the galaxy masses are theoretically good discriminators of
$\Omega_M$ given some observed MW and M31 separation and radial
velocity, current estimates of the masses are insufficient to make any
cosmological claims.  We then discuss additional complications to
expand this analysis.
\end{abstract}

\section{Introduction}

The dynamics of the Local Group of galaxies (LG; see Van den Bergh
1999; 2000; Mateo 1998, and references therein for recent reviews) can
potentially provide a wealth of information both about the component
galaxies and the underlying cosmological parameters (Governato et
al. 1997).

Kahn \& Woltjer (1959; see Peebles 1993 for a discussion) are the
first to suggest that the radial velocity of M31 toward the Galaxy
could be used as a means of estimating the combined mass of the Local
Group system.  Since the separation and radial velocity of the two
galaxies were known, they reasoned that the Local Group could be
treated as a simple two-body problem.

The observed separation of the two galaxies is well constrained.
Holland (1998) estimates the separation to be 745 kpc, while Mateo
(1998) in his excellent review of the dwarf galaxies in the Local
group gives a distance of 770 kpc.  A middle value of 760 kpc is given
by van den Bergh (1999; 2000), a value we will adopt throughout the
remaining discussion.

\begin{figure}
\centerline{\psfig{figure=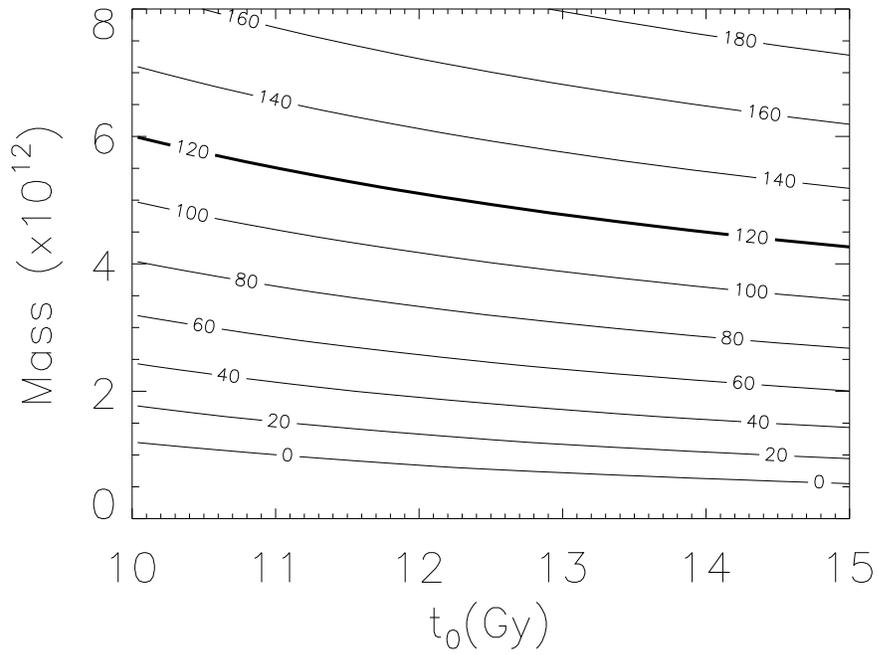,angle=0,height=4in}}
\caption{The relationship between imputed Local Group mass, the age of
the universe, and predicted radial velocity (shown in the contour
plot) according to the two-body approximation.  Note that under no
assumed age does the corresponding predicted mass match observations.}
\label{fg:2body}
\end{figure}

The cosmological parameters $H_0$, $\Omega_M$ and $\Omega_\Lambda$
uniquely give an age of the universe.  Thus the expected radial
velocity in this simplified model is a function of cosmology, observed
separation, and total mass of the system, as would be expected.  This
observed radial velocity, after subtracting the velocity of the Sun
around the center of the Milky Way is approximately -119 km/s (Binney
\& Tremaine, 1987 using solar velocity estimates of Lynden-Bell et
al., 1983).  We show the mass-velocity relation for two-body Local
Group models in Figure~\ref{fg:2body}, demonstrating that for
reasonable cosmologies, this method implies that the integrated Local
Group mass must be at least $4\times 10^{12} M_\odot$, and perhaps as
high as $6\times 10^{12} M_\odot$.

These estimated masses are on the high side of observations.  van den
Bergh (1999; 2000) gives an integrated mass of the Local Group of
$(2.3\pm 0.6)\times 10^{12} M\odot$, based on dispersion of component
galaxy radial velocity measurements.

Masses for the individual components have varied somewhat in recent
years.  Zaritsky (1999) estimates the mass of the Galaxy as $8.6\times
10^{11} M_\odot$ based on a combination of globular cluster
observations (out to $\sim 50$ kpc), and dwarf spheroidal observations
(out to $\sim 250$ kpc).  Wilkinson \& Evans(1999) primarily analyze
recent dwarf spheroidal measurements, and use them to compute a
detailed density profile of the Galaxy, yielding a much larger estimate
of $1.9\times 10^{12} M_\odot$, albeit with very large error bars.

Mass estimates for M31 are somewhat more tightly clustered, with
typical estimates of $2.3\times 10^{12} M_\odot$ estimated for the
entire subgroup (Courteau \& van den Bergh 1999), and $1.2\times
10^{12} M_\odot$, again based on radial velocity measurements of dwarf
spheroidals (Evans \& Wilkinson 2000).

A somewhat better treatment of the dynamics of the Galaxy-M31 system
can be had by treating the two galaxies explicitly as particles in the
cosmic flow.  This is the approach taken by Peebles (1989) in
describing a cosmological least action approach to solving particle
trajectories.

Treating M31, the Galaxy, six Local Group Dwarf galaxies, and a number
of nearby groups as point sources, he is able to calculate orbits
using a Least Action approach.  By varying cosmology, and comparing
the resulting distances and masses of the galaxies to their observed
values, he finds high Galaxy masses ($M_{LG}\simeq 5-6 \times 10^{12}
M_\odot$) for high $\Omega_M$ models, but good agreement with the
local distance scale.  Low $\Omega_M$ models, on the other hand,
yielded lower mass estimates ($M_{LG}=3.2-4.4\times 10^{12} M_\odot$),
and correspondingly worse distance estimates between M31 and the
Galaxy, but better distance estimates to nearby groups.  Since
different cosmologies provide better fits to different observables, he
makes no definitive conclusions about the true, underlying cosmology.

Shaya, Peebles, and Tully (1995) extend this technique considerably
beyond the Local Group, to model galaxy orbits out to 3000 km/s.  From
this survey, they estimate a value of $\Omega_M=0.17\pm0.10$.  Dunn
\& Laflamme (1993) find a similar result, also suggesting a low
$\Omega_M$.  Schmoldt \& Saha (1998) use the observed velocity, rather
than the distance, of M31 as a constraint, and measure a Local Group
mass of $4-8 \times 10^{12} h^{-1}M_\odot$, again, producing a much
larger mass than most observational evidence would suggest.

Branchini \& Carlberg (1995) object to this approach, however.  They
point out that in actual fact, galaxies did not exist as point masses
at any point in their history, and were, rather, collapsing halos
which overlap one another even today.  To test whether this would
affect corresponding mass estimates, they ran a number of N-body
simulations and compared the true masses and trajectories of
``galaxies'' in their simulations to the masses and trajectories which
could be imputed using a Least Action approach.  They find that the
point mass assumption is not necessarily a valid one, and that it
leads to an underestimated value of $\Omega_M$.  They suggest that
even $\Omega_M=1$ is consistent with Local Group dynamics.

Goldberg \& Spergel (2000, GS; Goldberg 2000) revisit the problem of
reconstructing orbits by introducing Perturbative Least Action (PLA).
PLA gets around the objections raised by Branchini and Carlberg by
treating galaxies not as point sources, but rather as part of a
continuous fluid which collapse over the course of a simulation.
Furthermore, since the orbits are constrained to grow linearly from a
uniform field at early times, the Zel'dovich (1970) approximation
necessarily holds.

Given PLA's ability to model the history of highly overdense systems,
the Local Group seems a natural target of investigation.  In
particular on may use radial velocity and distance measurements to
estimate the masses of M31 and the Galaxy, and to make predictions
about cosmology.  In \S~2, we review the Perturbative Least Action
method, and describe its applications.  In \S~3, we model a simplified
Local Group in several cosmologies, in essence duplicating the Kahn \&
Woltjer (1959) analysis for the case of known final masses and with
the assumption that galaxies form from the collapse of the initially
smooth density field.  We will show that even this toy model may
provide insight into the underlying cosmology.  Finally, in \S~4, we
discuss some potentially useful avenues of future study.

\section{Perturbative Least Action}
\label{sec:PLA}
 
Before describing the Local Group simulations, it might be productive
to briefly review the PLA method.  Interested readers will find a more
complete discussion in GS and Goldberg (2000).
 
The aim of PLA and methods like it is to reconstruct the initial
conditions of observed structure in the universe.  In essence, this
means that we wish to run an N-body calculation ``backwards in time.''
In order to accomplish this, we will make a number of simplifying
assumptions.  First, we treat the energy density of the universe as
consisting solely as Cold Dark Matter (CDM) and hence the only force
acting on the field will be gravity.  Secondly, the matter is treated
as a distribution of small point masses, which, when smoothed,
reproduce an observed density field.
 
In order to provide a context, let us imagine a particle field at
$a=0$ (where $a$ is the normalized expansion parameter of the
universe) which is uniformly distributed, and which has zero comoving
velocity.  We will label the unperturbed particle positions, $\{ {\bf
q}_i \}$ (where here and throughout, all particle positions are
assumed to be given in comoving coordinates).  At a slightly later
time, $t_i$, the particles are offset slightly, ${\bf p}_i(t_i)={\bf
p}({\bf q}_i, t_i)$, from their gridpoints, such that:
\begin{equation}
{\bf x}_i(t_i)={\bf q}_i+{\bf p}_i(t_i)\ ,
\label{eq:xrel}
\end{equation}

Linear perturbation theory can be used to demonstrate that small
perturbations grow monotonically with time.  In \S2.2.1, we showed
this yields the relation:
\begin{equation}
{\bf p}_i(t)=\frac{D(t)}{D(t_i)}{\bf p}_i(t_i)\ ,
\label{eq:prel}
\end{equation}
where $D(t)$ is the cosmologically determined linear growth factor,
normalized to unity at the present.  The combination of
equations~(\ref{eq:xrel}) and (\ref{eq:prel}) are known as the
Zel'dovich (1970) approximation, and describes the growth of structure
well so long as perturbations are small.

As perturbations grow to the point when particle trajectories cross
one another, the orbits become more complex.  This is essentially why
N-body simulations are needed in the first place; the trajectories of
particles are strongly coupled, and no analytic solutions exist for
orbits of more than two interacting particles.

In general, N-body simulations are run forward in time.  However,
Peebles (1989) pointed out that if we specify the initial and final
position of a particle rather than the initial position and velocity,
the trajectory may be solved by finding the path which minimizes the
action (the time integral of the Lagrangian):
\begin{equation}
S\equiv \sum_i \int_0^{t_0} dt \ {\cal L}_i= \sum_i \int dt \left( 
\frac{a^2\dot{\bf x}_{i}^2}{2}-\frac{\phi_i}{2} \right)\ ,
\label{eq:actdef}
\end{equation}
where $\phi_i$ is the potential on particle $i$.

While Peebles (1989; 1993; 1994) and others (Shaya, Peebles, \& Tully
1995; Schmoldt \& Saha 1998) have historically calculated the particle
potentials ($\phi_i$) by direct summation, this is impractical as the
number of particles gets large.  We get around this by using a
Particle Mesh (PM) Poisson solver (Hockney \& Eastwood 1981), which
requires computing time of order $N{\rm log}_2N$, rather than $N^2$.
Nusser and Branchini (1999) use a similar approach in their Least
Action technique by utilizing a tree code for their Poisson solver.

The PLA method involves running a randomly realized N-body simulation,
and storing the output trajectories which we will call ${\bf
x}^{(0)}_i(t)$.  Since the ``unperturbed'' orbit is known to satisfy
the equations of motion at all times, we may express the ``perturbed''
trajectory as:
\begin{equation}
{\bf x}_i(t)={\bf x}_i^{(0)}(t)+{\bf x}_i^{(1)}(t)\ .
\label{eq:perturb}
\end{equation}

These trajectories are then subject to the constraints ${\bf
x}_i(0)={\bf q}_i$ and ${\bf x}_i(t_0)={\bf x}_i^F$.  GS and Goldberg
(2000) discuss how one generates the final the set of final
constraints, ${\bf x}_i^F$ for a given set of observations.  

Once we've computed the boundary constraints on each particle, it
remains for the orbit of each particle to be parameterized so that we
can find the minimal action orbits. We express the trajectory of each
particle as:
\begin{equation}
{\bf x}_i(t)={\bf x}_i^{(0)}(t)+D(t){\bf x}^{(1)F}_i+
\sum_{n=1}^{n_{max}} {\bf C}_{i,n}f_n(t) \ ,
\label{eq:path}
\end{equation}
where ${\bf x}_i^{(1)F}\equiv{\bf x}_i^F-{\bf x}_i^{(1)}(t_0)$, ${\bf
C}_{i,n}$ are a set of coefficients, and $f_n(t)$, are a set of basis
functions.  We have found it useful to use the form:
\begin{equation}
f_{n}(t)=\sum_{m=n}^{n_{max}}b_{mn}D(t)\left[1-D(t)^m\right]\ .
\end{equation}
As a result, only the first basis function varies as slowly
as linearly at early times, and no decaying modes can be expressed.
Thus, we are implicitly assuming that the primordial fluctuations are
purely growing modes.

Using the form of the trajectory in equation~(\ref{eq:path}), we may
express the derivatives of the action with respect to the coefficients
as:
\begin{equation}
\frac{\partial S}{\partial {\bf C}_{in}}= \int dt \left[
\dot{f}_{n}(t)a^{2}\dot{\bf x}^{(1)}+f_{n}\left(
\nabla\phi^{(0)}_i-\nabla\phi_i \right) \right]\ ,
\label{eq:PLA}
\end{equation}
where $\phi_i^{(0)}$, is the potential on particle, $i$, in the
unperturbed potential field.  

An integral by parts of equation~(\ref{eq:PLA}) yields the
expression:
\begin{equation}
\frac{\partial S}{\partial {\bf C}_{i,n}} =\int dt f_n \left[
-\frac{\partial (a^2\dot{\bf x}_i^{(1)})}{\partial t} +\left(
\nabla\phi^{(0)}_i-\nabla\phi_i \right) \right]=0\ .
\end{equation}
Since the expression in the brackets is always satisfied whenever the
equations of motion of the individual particles are satisfied, we may
equivalently minimize:
\begin{equation}
X^2=\sum_i \int dt |W(t) {\bf g}_i(t) |^2
\label{eq:Chi_eq}
\end{equation}
where $W(t)$ is an arbitrary weighting function and
\begin{equation}
{\bf g}_i(t)\equiv -\frac{\partial (a^2\dot{\bf x}_i^{(1)})}{\partial t}
+\left( \nabla\phi^{(0)}_i-\nabla\phi_i \right)
\end{equation}
By finding values of the coefficients for which
equation~(\ref{eq:Chi_eq}) vanishes, we find trajectories which
satisfy the equations of motion.  We then evaluate the trajectories at
early times, and use the positions and velocities at high redshift as
input into an ordinary N-body code.  The output of the
N-body code may be further used as input to a new iteration for PLA.

\section{A Two Halo Model of an Isolated Local Group}
\label{sec:twohalo}

\subsection{The Simulations}

We now turn our attention to the simulations used to reconstruct
the Local Group.  Our primary goal in these simulations will be to
identify the radial velocity between the two principle elements of the
Local Group (the Galaxy and M31), as a function of galaxy mass and
cosmology.

In order to do this, we ran two sets of simulations, one with
$\Omega_M=0.3$, and $\Omega_\Lambda=0.7$ (Cosmology 1), and the other,
with $\Omega_M=1$ and $\Omega_\Lambda=0$ (Cosmology 2).  In each case,
we assumed $h=0.65$, where the Hubble constant, $H_0$, is define as
$100h$ km/s/Mpc.  The PLA reconstructions were run in a $(10 {\rm
Mpc})^3$ box, with $64^3$ particles, and $128^3$ gridcells, resulting
in a separation of $\sim 10$ gridcells between the centers of the two
galaxies.

For each cosmology, we set up 5 different target particle fields, each
based on somewhat differing assumed galaxy masses.  The density
profiles of the target used the form of the halo density given by
Dehnen and Binney (1998):
\begin{equation}
\rho_h=\rho_0\left(\frac{r}{r_0}\right)^{-\gamma}\left(1+\frac{r}{r_0}\right)^{\gamma-\beta}
\exp (-r^2/r_t^2)\ ,
\end{equation}
where we have explicitly assumed isotropic halos, and $\gamma$,
$\beta$, $r_0$, $\rho_0$, and $r_t$ are parameters of the model.  We
chose this parameterization over the more familiar form of Navarro,
Frenk, \& White (1997) because Dehnen \& Binney provide fits for the
Galaxy according to this form.

For each model, we hold $\gamma=-2$, $\beta=2.959$, $r_0=3.83$ kpc,
and $r_t=700$ kpc fixed, and vary values of $\rho_0$ in order to probe
different combinations of galaxy masses.  We find variations in mass
and cosmology were clearly more important than the fine details of
halo density profile.  Moreover, our resolution scale was such that we
treat the MW ``galaxy'' halo as including all of its nearby satellite
galaxies.

\subsection{Results}

A summary of the results of the ten realizations are shown in
Table~\ref{tab:results}.  In it, we show the masses of the
reconstructed galaxies within $\overline\delta(r)=200$, their
velocities along the x-axis, and the masses of the ``target'' galaxies
within $\overline\delta(r)=200$.  Velocities are given in km/s with
respect to expanding coordinate system.  Since the assumed separation
of the galaxies is $760 \ {\rm kpc}$, and $h=0.65$, the ``observed''
radial velocity for each is given as:
\begin{equation}
u_r=v_r^{M31}-v_r^{MW}+H_0r=v_r^{M31}-v_r^{MW}+49\ .
\label{eq:velrel}
\end{equation}

Overall, the quality of the reconstruction is quite good.  Comparison
of the observed galaxy masses to the masses measured in the target
field yields agreement to about $20\%$.  Generally, the reconstructed
mass is somewhat smaller than the target mass since small errors in
the computed particle trajectories result in an overall smoothing in
the density field.  The axial symmetry of the problem resulted in
tangential velocities of typically only 1 or 2 km/s.

We now turn to an interpretation of these results.
Figure~\ref{fg:massvel} shows a plot of the data in
Table~\ref{tab:results}, combined with the equation~(\ref{eq:velrel}).
What is quite intriguing about these mass estimates is that according
to the PLA reconstruction, the two-body mass estimate of the LG {\it
over}estimates the mass of the galaxy halos.  However, one's intuition
would suggest that since the two-body estimate does not evolve from
collapsed halos, the true forces should be softer than those used by
Kahn \& Woltjer.  These means that we might expect that the two-body
estimate underestimates the true mass.  However, this argument may be
flawed as it does not take into account the detailed collapse of the
galaxies, which makes the situation much more complex.  Moreover,
Branchini \& Carlberg (1995) suggest the same effect using N-body
simulations.

\begin{figure}
\centerline{\psfig{figure=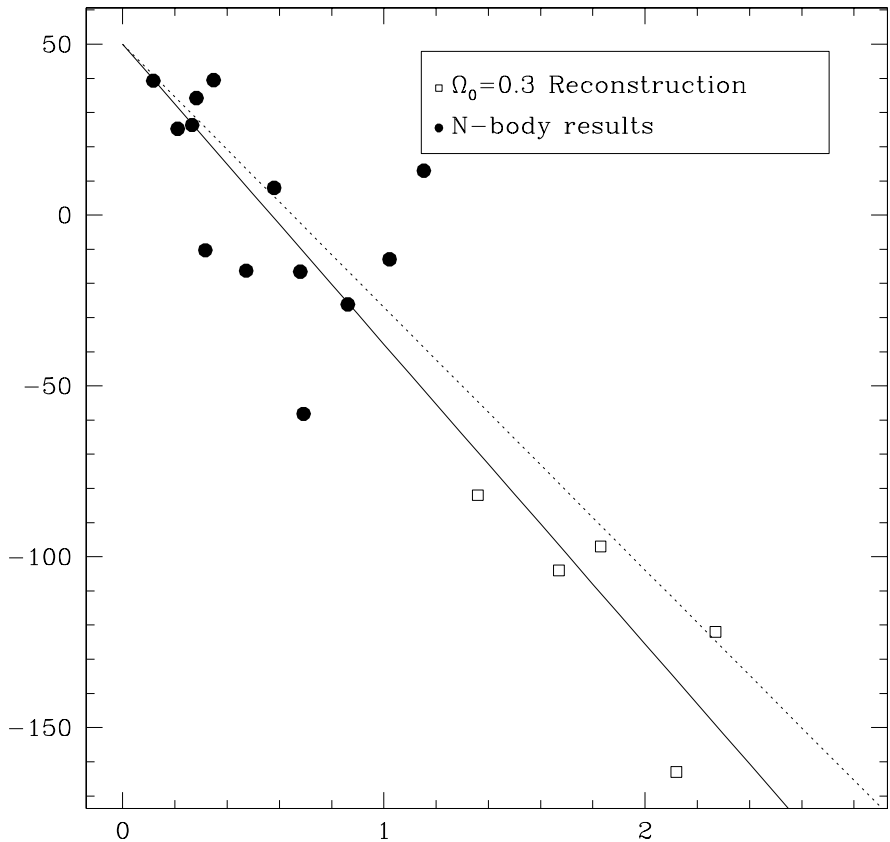,angle=0,height=5in}}
\caption{As in the previous figure, the relationship between total
Local Group mass and ``observed'' radial velocities for the
reconstructed LG models under the $\Omega_M=0.3$ PLA reconstruction,
and a randomly realized set of Local Groups found in an N-body
simulation.  The open squares indicate the PLA reconstruction, and
the filled circles show the simulation.  The solid line represents the
best fit mass-velocity relation for PLA, while the dashed line
represents a similar fit for the simulations. }
\label{fg:blobs}
\end{figure}

We perform a similar experiment here in order to check the consistency
of the mass and velocity estimates.  We ran a series of N-body
simulations in a flat, $\Omega_M=0.3$ cosmology, and extracted out all
halos within a mean overdensity of $\overline\delta=200$.  We then
selected those halo pairs which formed two member associated groups
(i.e. those which were falling in toward one another).  N-body codes
all for a self-consistent scaling of the box size, velocities, and
masses of the particles.  For each pair, we set the scaling factor to
give a galaxy separation of $760$ kpc.  The resulting group masses and
velocities are plotted in Figure~\ref{fg:blobs} along with the masses
and velocities for the reconstructed LG in the $\Omega_M=0.3$
reconstruction set.

It is clear from Figure~\ref{fg:blobs} that the overall mass-velocity
relation determined from the reconstruction is generally consistent
with that generated from randomly realized N-body simulations, though
the best fit slope of this relation differs by $\sim 12\%$ (suggesting
that the PLA reconstruction provides a slight mass underestimate).
Moreover, it is clear that there is significantly greater scatter in
the random realizations than in the reconstructed halos.  This is due
to the fact that the PLA reconstruction contains no outside, tidal
forces, while the randomly realized field will necessarily have small
halos scattered nearby.

As for the comparison of the mass-velocity relationship between the
two cosmologies in the reconstruction, as illustrated in
Figure~\ref{fg:massvel}, Cosmology 1 produces a systematically larger
radial velocity for a given Local Group mass than does Cosmology 2.
This accords with the relationship given by the two-body problem, and
shown in Figure~\ref{fg:2body}, but we may use a slightly better
argument in order to estimate what this scaling ought to be.

We will address this scaling using a simple Zel'dovich model.
Linear theory predicts that for small perturbations, the velocity
field can be expressed as:
\begin{equation}
\nabla \cdot {\bf v}({\bf x},t) =
-\frac{a\dot{D}}{\dot{a}D}\delta({\bf x},t)
=-f(\Omega_M,\Omega_\Lambda,t)\delta({\bf x},t)\ .
\end{equation}
However, $f(\Omega_M,\Omega_\Lambda,t_0) \propto \Omega^{0.6}$, as was
pointed out by Peebles (1980, \S 14).  In order to form a structure of
a given mass, $\delta\propto M/\Omega_M$.  Thus, we find the relation:
\begin{equation}
{\bf v} -\propto \Omega_{m}^{-0.4}M \ ,
\label{eq:zeldapprox}
\end{equation}
where the constant of proportionality is assumed to be a function of
separation and Hubble constant, both of which are kept constant in all
of the simulations.

In fact, we find a somewhat shallower relation, as illustrated by the
best fit lines in Figure~\ref{fg:massvel}.  If we assume:
\begin{equation}
v_r=u_r+H_0 r\propto M_{tot}\Omega_M^{n}\ ,
\end{equation}
then the observed
coefficient for our simulations is only $n\simeq -0.27$.

\begin{figure}
\centerline{\psfig{figure=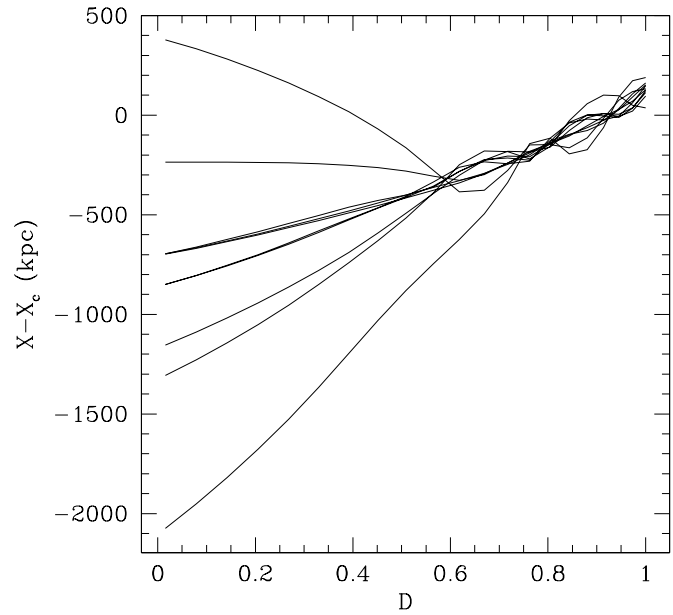,angle=0,height=4in}}
\caption{X-coordinate orbits of 10 randomly selected MW particles in
Simulation 5 of Cosmology 1 ($\Omega_M=0.3$).  Each particle was
selected from $100-200$ kpc of the MW center.  Note that there are
typically several orbit crossings over the course of the
realization. }
\label{fg:xpos}
\end{figure}

\begin{figure}
\centerline{\psfig{figure=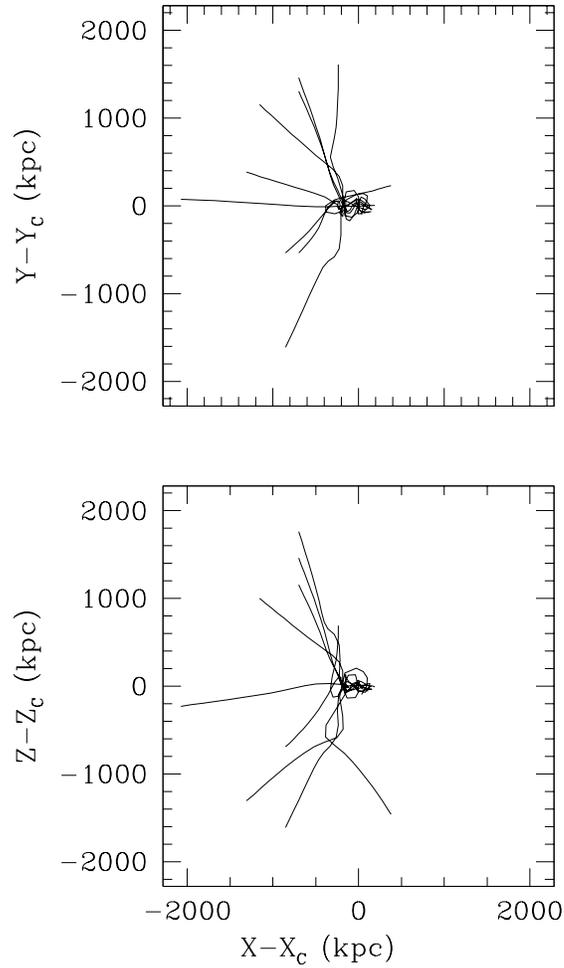,angle=0,height=5.5in}}
\caption{A trace of 10 orbits randomly selected MW particles in
Simulation 5 of Cosmology 1 ($\Omega_M=0.3$).  The top panel shows the
orbits in the X-Z plane, while the bottom panel shows the orbits in
the X-Y plane.}
\label{fg:orbits}
\end{figure}

In Figures~\ref{fg:xpos} and \ref{fg:orbits} we show some the
orbits of randomly selected MW particles in Model 5 of Cosmology 1
($\Omega_M=0.3$) in order to illustrate that even though the initial
and final particle positions were constrained to provide a first
collapse solution, this does not mean that the particles have not
experienced orbit crossings.  Figure~\ref{fg:xpos} shows this
explicitly, by illustrating the x-coordinate (in kpc) of 10 randomly
selected MW particles.  Each particle selected was found between
$100-200$ kpc from the Galaxy center at $z=0$.  As expected from a
typical orbital velocity of $100-200$ km/s, they all seem to have
orbited the galaxy center of mass several times by the end of the
simulation.  This is likewise illustrated by
Figure~\ref{fg:orbits}, which shows a trace of the orbits as
viewed from the z- and y- axes.

As another interesting test of the robustness of the reconstructions, we
took one of the simulations, (\# 5 in Cosmology 1), and smoothed the
small scale power using a Gaussian filter with diameter 400 kpc.  We
then increased the resolution by a factor of two, as described in
GS, multiplying the number of particles by 8, and filling in
small scale power randomly using an assumed CDM power spectrum.  We
then ran that resulting simulation through a PM code, and recovered
the LG characteristics, as shown in Model 5a in Table
~\ref{tab:results}.  While the total radial velocity did not
change much from the low to high-resolution form of the
reconstruction, and while the mass of ``M31'' stayed approximately
constant, The Galaxy increased by $\sim 70\%$ in mass in the high
resolution case.

Though the specific details of the reconstruction differ somewhat in
increasing the resolution, it is reassuring to see that the Local
Group reconstruction is qualitatively similar to the low resolution
case. 

\section{Discussion}

We have endeavored to reproduce the two major systems in the Local
Group, the Milky Way, and M31 using the PLA approach.  This was
motivated by the seeming discrepancy introduced by the masses
estimated by using the timing argument, and the much lower masses
given by observations of the dynamics around the two halos.  We have
shown that by treating the galaxies as halos, rather than as point
sources, and constraining the current separation and peculiar
velocities to those observed, a much more consistent estimate of
masses may be obtained.

\begin{figure}
\centerline{\psfig{figure=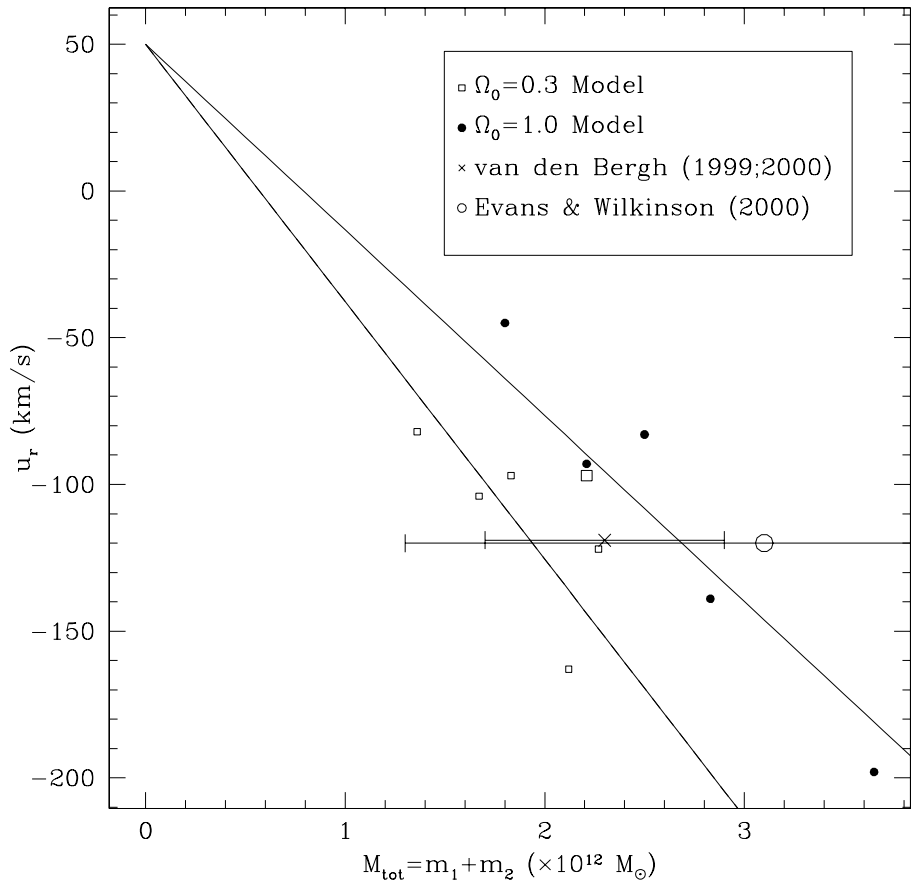,angle=0,height=5in}}
\caption{The relationship between total Local Group mass and
``observed'' radial velocities for the reconstructed LG models under
$\Omega_M=0.3$ and $\Omega_M=1.0$ cosmologies, with a comparison to
actual mass estimates of the Local Group.  The open squares indicate
Cosmology 1 ($\Omega_M=0.3$), while the larger square shows the
results of the high-resolution realization.  The filled circles
represent Cosmology 2 ($\Omega_M=1.0$).  Note that given the error bars
in the observed masses of M31 and the Milky way, strong constraints
cannot yet be made on cosmology from these models.}
\label{fg:massvel}
\end{figure}

As illustrated in Figure~\ref{fg:massvel}, though PLA can be
applied to the Local Group to theoretically discriminate between a
high- and low-density universe, the mass observations are currently
not sufficiently tight to make such a determination practical with
only two galaxies.  However, it may be practical to re-address the
specific program suggested by Shaya, Peebles, \& Tully (1995).  In
that, we do not simply treat the Local Group as two objects, but
rather, treat the two objects in a much larger, and interacting
context.  

In future work, we will reconstruct not only the major halos of the
Local Group, but also add the effects of nearby groups like the
Sculptor Group (Jerjen, Freeman, \& Bingelli, 1998), the Centaurus A
Group (van den Bergh 1999a), and the Antlia-Sextans group (van den
Bergh 2000a).  Indeed, to properly treat the problem, one would prefer
an external potential field out to distances well into the linear
regime.  While this last may be applied directly using linear theory,
the nearby groups must be themselves reconstructed.  This requires a
larger simulation volume, but the resolution length must be maintained
in order to differentiate between the Milky Way and M31.  Thus, the
simulation scale, (e.g., the number of particles) and thus its
computation time and memory requirements, will get large with the
application of even a few nearby groups, as the distance to even the
nearest group, Ant-Sex, is about 3 times the radius of the Local
Group, itself.  A realistic reconstruction would thus require a
simulation box of about three times the size of those used in the
present discussion.

\acknowledgements

I would like to gratefully acknowledge many helpful comments by David
Spergel, Michael Strauss, and Jim Peebles.  This work was supported by
an NSF Graduate Research Fellowship and NASA ATP grant NAG5-7154.

\newpage

\begin{table}
\begin{tabular}{|c||c|c|c|c|c|c|}
\hline
\multicolumn{7}{|c|}{Cosmology 1: $\Omega_M=0.3$} \\ \hline
Model & $M_{MW} (\times 10^{12} M_\odot)$ &  $M_{M31} (\times 10^{12}
M_\odot)$ & $v_r^{MW}$ & $v_r^{M31}$ & $M_{MW}^{Target}$ & $M_{M31}^{Target}$\\
\hline\hline
1 & 1.41 & 0.86 & 69 &  -103 &1.91 & 1.00  \\ \hline
2 & 0.89 & 0.94 & 75 & -72  & 1.28 & 1.31 \\ \hline
3 & 1.02 & 1.10 & 106 &  -107 & 1.38 & 1.43 \\ \hline
4 & 0.62 & 0.74 & 67 & -65 & 1.14 & 1.18 \\ \hline
5 & 0.86 & 0.81 & 75 &  -79 & 1.49 & 1.27 \\ \hline
5a & 1.48 & 0.73 & 48 & -99 & 1.49 & 1.27 \\ \hline\hline
\multicolumn{7}{|c|}{Cosmology 2: $\Omega_M=1.0$} \\ \hline
 & $M_{MW} (\times 10^{12} M_\odot)$ &  $M_{M31} (\times 10^{12}
M_\odot)$
& $v_r^{MW}$ & $v_r^{M31}$  & $M_{MW}^{Target}$ & $M_{M31}^{Target}$
\\ \hline\hline
1 & 0.99 & 1.02 & 46 &  -49 & 1.02 & 0.87 \\ \hline
2 & 1.14 & 1.07 & 73 & -70  & 1.22 & 1.08 \\ \hline
3 & 1.49 & 1.01 & 53 &  -80  & 1.56 & 1.12 \\ \hline
4 & 2.09 & 1.56 & 100 & -148  & 2.90 & 1.78 \\ \hline
5 & 1.74 & 1.09 & 79 &  -110  & 2.25 & 1.53 \\ \hline
\end{tabular}
\caption{A summary of the results of the various Local Group
reconstructions.  Listed are the integrated masses out to an
overdensity of $\delta=200$, in units of $10^{12} M_\odot$, the mean
velocity along the x-axis of the particles within each galaxy in units
of km/s, and the actual mass of the galaxies used in the target
density field.  Note that to get the observed radial velocity, one
needs to compute $u_r=v_{M31}-v_{MW}+H_0r$, where the last term is
approximately $49$ km/s for $h=0.65$ and $r=760 kpc$.}
\label{tab:results}
\end{table}

\end{document}